\documentstyle[graphics,12pt]{article}
\newcommand{\myvec}[1]{\displaystyle {\bf #1}}
\begin{document}
\title{Investigation of \protect $^{4}He_{3}\protect $ trimer on the base of Faddeev
equations in configuration space.}
\author{V.Roudnev\footnote{e-mail: roudnev@cph10.phys.spbu.ru}, S.Yakovlev 
\vspace{4mm} \\
Institute for Physics, \\ St.Petersburg State University,\\ Russia}
\date{ }
\maketitle
\begin{abstract}
Precise numerical calculations of bound states of three-atomic Helium cluster are performed.
The modern techniques of solution of Faddeev equations are combined to obtain
an efficient numerical scheme. Binding energies and other observables for ground
and excited states are calculated. Geometric properties of the clusters are discussed. 
\end{abstract}
\section{Introduction}
Small clusters of Helium attract the attention of specialists in different fields
of physics. Fine experimental techniques are developed to observe these
clusters \cite{Experiment,Experiment2,Experiment3}. Different 
quantum chemistry approaches are used to produce numerous potential models
of He-He interaction \cite{LM,Az79,Az87,Az91,VanMourik,TTY}.
Model-free Monte-Carlo calculations were performed to check the accuracy 
of the models \cite{MonteCarlo}.
The special attention is payed to three-body Helium clusters 
\cite{Lim1,Lim2,Gloeckle,Pandharipande,Moto1,Carbonell,Fedorov} because of
their possible near-Efimov behavior \cite{Efimov}. 
Complicated shape of the model potentials also makes Helium trimer a perfect 
touchstone for the computational methods of three-body bound state 
calculations \cite{CPC}.

Although the investigation of He$_{3}$ lasts already more than 20 years
\cite{Lim1}, some important physical questions has not received definite answer yet.
One of the questions is dealt with speculations on Efimov-like nature of the
He$_{3}$ bound states: how many excited states are supported by the
best known model interactions? Can one estimate any differences in the number
of bound states varying the model potentials being limited by the accuracy of
contemporary models? Another important question is dealt with the characteristics
of He$_{3}$ bound states. Can He$_3$ trimers influence
the results of experimental measurement of He$_{2}$ dimer characteristics?
To answer these questions one should know such important
characteristics of the He$_{3}$ cluster as the mean square radius of different states 
of the trimer and its geometric shape.

In this paper we investigate the $^4$He trimer performing direct calculations
of $^{4}$He$_{3}$ bound states with different He-He model potentials.
We base our calculations on Faddeev equations in configuration space because of 
simplicity of numerical approximation of Faddeev components in comparison 
with a wave function. In the case of Faddeev equations the boundary conditions 
are also much simpler.

In the following sections the equations we have solved numerically are described,
some observables for different states of He$_3$ and for He$_2$ are presented.

\section{Faddeev equations for bound states}
According to the Faddeev formalism \cite{FaddMerk} the wave function of three particles
is expressed in terms of Faddeev components $\Phi$
\begin{equation}
\label{WFDef}
  \Psi( \myvec {x_1},\myvec{y_1})= 	\Phi_1 ( \myvec {x_1},\myvec{y_1})+
					\Phi_2 ( \myvec {x_2},\myvec{y_2})+
			       		\Phi_3 ( \myvec {x_3},\myvec{y_3}) \: ,
\end{equation}
where $\myvec {x}_{\alpha }$ and $\myvec {y}_{\alpha }$ are Jackobi coordinates 
corresponding to the fixed pair $\alpha$
\begin{equation}
\begin{array}{c}
 \myvec {x}_{\alpha }=(\frac{2m_{\beta }m_{\gamma }}{m_{\beta }+m_{\gamma
 }})^{\frac{1}{2}}(\myvec {r}_{\beta }-\myvec {r}_{\gamma })\ ,\\
 \myvec {y}_{\alpha }=(\frac{2m_{\alpha }(m_{\beta }+m_{\gamma })}{m_{\alpha
 }+m_{\beta }+m_{\gamma }})^{\frac{1}{2}}(\myvec {r}_{\alpha }-\frac{m_{\beta
 }\myvec {r}_{\beta }+m_{\gamma }\myvec {r}_{\gamma }}{m_{\beta }+m_{\gamma }})
 \ .
\end{array}
\label{Jacoord}
\end{equation}
Here $\myvec {r}_{\alpha }$ are the positions of the particles in the center-of-mass
frame. The Faddeev components obey the set of three equations
\begin{equation}
\label{eqf}
\begin{array}{c}
 (-\Delta _{x}-\Delta _{y}+V_{\alpha }(\myvec{x}_{\alpha })-E)\Phi _{\alpha }(\myvec {x}_{\alpha },\myvec {y}_{\alpha })=-V_{\alpha }(x_{\alpha })\sum _{\beta \neq \alpha }\Phi _{\beta }(\myvec {x}_{\beta },\myvec {y}_{\beta })
  \\ 
  \alpha=1,2,3  
\end{array} \, ,
\end{equation}
where $V_{\alpha }(\myvec{x}_{\alpha })$ stands for pairwise potential.
To make this system of equations suitable for numerical calculations
one should take into account the symmetries of the physical system. 
Exploiting the identity of Helium atoms in the trimer one can reduce the equations (\ref{eqf}) 
to one equation \cite{FaddMerk}. 
Since all the model potentials are central it is possible to factor out the degrees of freedom corresponding to
the rotations of the whole cluster \cite{TAM}. For the case of zero total angular momentum 
the reduced Faddeev equation reads
\begin{equation}
\label{Fadd3}
\begin{array}{rl} \displaystyle
  (-\frac{\partial ^{2}}{\partial x^{2}} & \displaystyle
   -\frac{\partial ^{2}}{\partial y^{2}}
   -(\frac{1}{x^{2}}+\frac{1}{y^{2}})
   \frac{\partial }{\partial z}(1-z^{2})^{\frac{1}{2}}
   \frac{\partial }{\partial z}+ \\  \displaystyle
   & \displaystyle +xyV(x)(1+C^{+}+C^{-})\frac{1}{xy}-E)\Phi (x,y,z)=0\: .
\end{array}
\end{equation}
Here 
\begin{equation}
\label{intrcoord}
\begin{array}{c}
  x=|\myvec {x}|\, ,\\
  y=|\myvec {y}|\, ,\\ 
  \displaystyle 
  z=\frac{(\myvec {x},\myvec {y})}{xy} \, ,
\end{array}
\end{equation}
$C^{+}$ and $C^{-}$ are cyclic and anticyclic permutation operators
acting on the coordinates $x$, $y$ and $z$ as follows
\[
\begin{array}{l} \displaystyle
C^{\pm }x=(\frac{x^{2}}{4}+\frac{3y^{2}}{4}\mp
\frac{\sqrt{3}}{2}xyz)^{1/2} \; ,\\  \displaystyle
C^{\pm }y=(\frac{3x^{2}}{4}+\frac{y^{2}}{4}\pm
\frac{\sqrt{3}}{2}xyz)^{1/2}\; ,\\  \displaystyle
C^{\pm }z=\frac{\pm \frac{\sqrt{3}x^{2}}{4}\mp
\frac{\sqrt{3}y^{2}}{4}-\frac{1}{2}xyz}{C^{\pm }x\, C^{\pm }y}\; .
\end{array}
\]

The asymptotic boundary condition for bound states consists of two terms \cite{FaddMerk}
\[
 \Phi (x,y,z)\sim \: \phi _{2}(x)e^{-k_{y}y}+A(\frac{x}{y},z)\frac{e^{-k_{3}(x^{2}+y^{2})^{\frac{1}{2}}}}{(x^{2}+y^{2})^{\frac{1}{4}}}\: ,
\]
where $\phi _{2}(x)$ is the two-body bound state wave function, $k_{y}=\sqrt{E_{2}-E_{3}}$,
$k_{3}=\sqrt{-E_{3}}$, $E_{2}$ is the energy of the two-body bound
state and $E_{3}$ is the energy of the three-body system. The second
term corresponding to virtual decay of three body bound state into three single particles
decreases much faster than the first one which corresponds to virtual decay into
a particle and two-body cluster. In our calculations we neglect the
second term in the asymptotics introducing the following approximate boundary conditions for the Faddeev
component at sufficiently large distances $R_{x}$ and $R_{y}$ 
\begin{equation}
\label{bc}
\begin{array}{l}  \displaystyle
  \frac{\partial _{x}\Phi (x,y,z)\lfloor_{x=R_x}}
  {\Phi (x,y,z)\lfloor_{x=R_x}}=k_2\equiv
                     i\sqrt{E_{2}}\: , \\  \displaystyle
  \frac{\partial _{_{y}}\Phi (x,y,z)\lfloor_{y=R_y}}
       {\Phi (x,y,z)\lfloor_{y=R_y}}=k_{y} \: .
\end{array}
\end{equation}

To calculate the bound state energy and the corresponding Faddeev component
one has to solve the equation (\ref{Fadd3}) with the approximate boundary condition
(\ref{bc}). The numerical scheme we have chosen to perform the calculations
is based on tensor-trick algorithm \cite{Groning}. In this paper we do not
describe the realization of the numerical methods exploited but only underline
some essential features of our approach. They are
\begin{enumerate}
\item total angular momentum representation \cite{TAM},
\item tensor-trick algorithm \cite{Groning},
\item Cartesian coordinates \cite{Carbonell}.
\end{enumerate}
The total angular momentum representation itself is a strong method of partial
analysis allowing to take into account contribution of all the angular momentum
states of two-body subsystems at once \cite{TAM}. Tensor-trick
algorithm \cite{Groning} is known to be a powerful method of solution of Faddeev
equations for bound states. Being applied to the equations in total angular momentum 
representation it leads to effective computational scheme which makes possible to
use all the advantages of Cartesian coordinates. In particular
using Cartesian coordinates \cite{Carbonell} one can obtain a criterion to select 
the optimal grid in the coordinate $x$. This criterion comes from the asymptotic behavior
of Faddeev component
\[
\Phi (x,y,z)\sim \varphi _{2}(x)e^{-k_{y}y}\, ,
\]
where $\varphi _{2}(x)$ is the two-body bound state wave function. That
is why the properly chosen grid in $x$ should support the correct two-body
wave function. Comparing the binding energy of two-body subsystem calculated
on the "three-body" grid with the exact results one can estimate
the lower bound for a numerical error of a three-body calculation. 

Thus the usage of total angular momentum representation has allowed us to construct
an efficient numerical scheme combining the most advanced methods proposed during
the last decade.

\section{Results of calculations}
Having the equation (\ref{Fadd3}) solved numerically one has the value of
the energy of 3-body state $E_{3}$ and the corresponding Faddeev component
$\Phi (x,y,z)$ for a particular model potential. Comparing the observables
calculated for different potential models one can estimate the bounds limiting
the values of these observables for the real system. Eight different potential
models were used in our calculations: HFD-HE2 \cite{Az79}, HFD-B(He)
\cite{Az87}, LM2M1~\cite{Az91}, LM2M2~\cite{Az91}, HFDID~\cite{Az91},
LM2M1 and LM2M2 without add-on correction term \cite{Az91}, TTYPT~\cite{TTY}.
In the Tab.~1 we give the values of trimer energies for ground and excited
states. To confirm the accuracy of our calculation we also present the values
of dimer binding energy calculated on the grid used in three-body
calculations $\tilde{E_2}$ and the exact results $E_2$. The
difference between these values can be regarded as the lower bound for the
error of our approach. In the Tabs.~2 and~3 we demonstrate 
the convergence of the calculated energies with respect to the number
of grid points used in the calculations. The results of other authors for
the most known potentials are given in the Tab.~7. The best agreement
is observed with the results of \cite{Pandharipande} and \cite{Fedorov}. In
the ref. \cite{Pandharipande} no angular momentum cut-off is made, that makes
it the closest one to our approach. In all other papers some kind of partial
wave decomposition is performed and finite number of angular basic functions
is taken into account. The most complete basis is used in the ref. \cite{Fedorov}.
The agreement between our calculations and the result of \cite{Fedorov} for the excited state 
is impressive, but the ground state energy of \cite{Fedorov} is about one percent
less than our result. Consideration of the geometric properties of Helium trimer
can clarify the possible nature of this difference.

Since the Faddeev component is calculated, the wave function can be recovered
as follows
\[
  \psi (x,y,z)=\Phi (x,y,z)+xy\, (\frac{\Phi (x^{+},y^{+},z^{+})}{x^{+}y^{+}}+\frac{\Phi (x^{-},y^{-},z^{-})}{x^{-}y^{-}}) \ ,
\]
where $x^{\pm }=C^{\pm }x$ and $y^{\pm }=C^{\pm }y$. 
Having the wave function recovered one can investigate the shape properties
of the system. The most intuitive way to visualize the results of the calculations
is to draw a one-particle density function defined as
\[
  \rho (\myvec {r_{1}})=\int d\myvec {r_{2}}d\myvec {r_{3}}|\Psi (\myvec {r_{1}},\myvec {r_{2}},\myvec {r_{3}})|^{2}\, ,
\]
where 
\[
\Psi(\myvec {r_{1}},\myvec {r_{2}},\myvec {r_{3}})
   =
   \displaystyle 
   \frac{
     \psi(x(\myvec {r_{1}},\myvec {r_{2}},\myvec {r_{3}}),y(\myvec {r_{1}},\myvec {r_{2}},\myvec {r_{3}}),z(\myvec {r_{1}},\myvec {r_{2}},\myvec {r_{3}}))
     }{
       4\pi^3 x(\myvec {r_{1}},\myvec {r_{2}},\myvec {r_{3}}) y(\myvec {r_{1}},\myvec {r_{2}},\myvec {r_{3}})
       } \, ,
\]
the functions $x(\myvec {r_{1}},\myvec {r_{2}},\myvec {r_{3}})$,
$y(\myvec {r_{1}},\myvec {r_{2}},\myvec {r_{3}})$ and 
$z(\myvec {r_{1}},\myvec {r_{2}},\myvec {r_{3}})$ are defined according to (\ref{Jacoord}) and (\ref{intrcoord}), 
the function $\psi(x,y,z)$ is normalized to unit.
Due to the symmetry of the system the one-particle density function is a function
of the $r_{1}=|\myvec {r_{1}}|$ coordinate only. Taking into account the
relation $\myvec {r_{1}}=\sqrt{3}\myvec {y_{1}}$ we get
\[
  \rho (r)=\sqrt{3}\int dx\, dz\, |\psi (x,\frac{r}{\sqrt{3}},z)|^{2}\: .
\]
Omitting the integration over $z$ we define a conditional density function
$\rho (r,z)$ that presents a spatial distribution for the particle 1 when
the other two particles are located along the fixed axis. It is useful
to plot this function in coordinates $(r_l,r_a)$ such that $r_l=r z$ is
a projection of the particle 1 position to the axis connecting the other particles and 
$r_a=\frac{z}{|z|} r(1-z^2)^\frac{1}{2}$ is a projection to the orthogonal axis. Three-dimensional
plots of the function $\rho ((r_l^2+r_a^2)^{1/2},\cos \arctan \frac{r_l}{r_a})$ 
corresponding to the ground and excited
states of the trimer calculated with LM2M2 potential are presented on the Fig.~1 
and Fig.~2. The conditional density function of the ground state decreases
democratically in all the directions. The density function of the
excited state has two distinguishable maximums demonstrating the linear
structure of the cluster. This structure has a simple physical
explanation. The most probable positions of a particle in 
the excited state lie around two other particles. At the distances
where two particles are well separated the third one forms a dimer-like bound state
with each of them. This interpretation agrees with the clusterisation coefficients
presented in the Tab.~4. These coefficients are calculated as a norm of the
function $f_{c}$ defined as follows
\[
  f_{c}(y)=\int dx\, dz\, \Phi (x,y,z)\phi _{2}(x)\: ,
\]
where $\phi _{2}(x)$ is the dimer wave function. The values of $\|f_{c}(y)\|^{2}$
shown in the Tab.~5 demonstrate the dominating role of a two-body contribution
to the trimer excited state  whereas in the ground state this contribution
is rather small. We could suppose that this dominating contribution of the cluster
wave in the excited state has ensured fast convergence of the hyperspherical
adiabatic expansion in the paper \cite{Fedorov} to the correct value, but to
get the same order of accuracy for the ground state possibly more basic functions
should be taken into account. 

Very demonstrative example of the advantage of Faddeev equations over the Schroedinger
one in bound-state calculations is given in the Tabs.~8 and~9. Here we present
the  contribution of different angular states to the Faddeev component and to
the wave function calculated as
\[
\begin{array}{c}
  C_n=\|f_{n}(x,y)\|^2 \\ \displaystyle
  f_{n}(x,y)=\int ^{1}_{-1}dz\, F(x,y,z)P_{n}(z)\: ,
\end{array}
\]
where $P_{n}(z)$ are the normalized Legendre polynomials, $F(x,y,z)$
is the Faddeev component or the wave function, $n=0,2,4$. The angular coefficients
for the Faddeev component decrease much faster than the wave function coefficients.
The Tab.~8 also demonstrates that more angular functions should be taken into account
in the ground state calculations.

\section{Conclusions}
The high accuracy calculations of He$_3$ bound states were performed on the base 
of the most advanced few-body calculations techniques. Eight different potential models were used.
For every potential model, either more (LM2M2, TTYPT) or less realistic one (LM2M2a, HFD-ID), 
two and only two bound states are found.
The properties of these states are very different. The ground state is strongly bound, whereas
the binding energy of the excited state is comparable with the binding energy
of dimer. The sizes of these two states also differs much. The characteristic
size of the ground state either estimated by $\langle r \rangle$ or 
$\langle r^{2}\rangle^{1/2}$ (Tabs.~5 and~6) is approximately 10 times less 
than the size of dimer molecule, but the size
of the excited state has the same order of magnitude with the dimer's one. This estimation
shows the necessity to check for the absence of trimers in the experimental
media during the measurement of dimer properties and vice versa.

\section*{Acknowledgements}
One of the authors (VR) is grateful to the Leonhard Euler Program for
financial support. The authors are thankful to Freie Universit\"at Berlin
where the final stage of this work was performed. We are also thankful to
Robert Schrader for his warm hospitality during our visit to Berlin.

\newpage
\subsection*{List of tables}
\begin{enumerate}
\item The energy of the He\protect$_{2}\protect$and He\protect$_{3}\protect$
bound states
\item Convergence of the He\protect$_{3}\protect$excited state energy with respect
to the number of gridpoints
\item Convergence of the He\protect$_{3}\protect$ground state energy with respect
to the number of gridpoints
\item Contribution of cluster wave to the Faddeev component
\item The mean square radius of Helium molecules
\item The mean radius of Helium molecules
\item Comparison with the results of other authors
\item Contribution of different two-body angular states to the Faddeev component
\item Contribution of different two-body angular states to the wave function
\end{enumerate}
\subsection*{List of figures}
\begin{enumerate}
\item Conditional one-particle density function of the He$_3$ ground state 
\item Conditional one-particle density function of the He$_3$ excited state
\item He$_3$ ground state density function
\item He$_3$ excited state density function
\item He$_2$ density function
\end{enumerate}

\newpage 
\begin{table}
\caption{The energy of the He\protect$_{2}\protect$ and He\protect$_{3}\protect$
bound states}
\vspace{5mm}
{ \begin{tabular}{|c|c|c|c|c|}
\hline 
Potential&
$E_{2}$,mK & $\widetilde{E_{2}}$, mK & $E_{3}$, K & $E^{*}_{3}$, mK \\
\hline 
\hline 
HFD-A&-0.830124&-0.8305&0.11713&1.665\\
\hline 
HFD-B&-1.685419&-1.68540&0.13298&2.734\\
\hline 
HFD-ID&-0.40229&-0.4024&0.10612&1.06\\
\hline 
LM2M1&-1.20909&-1.212&0.12465&2.155\\
\hline 
LM2M2&-1.303482&-1.304&0.12641&2.271\\
\hline 
LM2M1a&-1.52590&-1.527&0.13024&2.543\\
\hline 
LM2M2a&-1.798436&-1.795&0.13471&2.868\\
\hline 
TTY&-1.312262&-1.3121&0.12640&2.280\\
\hline 
\end{tabular}\par}
\end{table}  
\begin{table}
\caption{Convergence of the He\protect$_{3}\protect$ excited state energy with respect
to the number of gridpoints}
\vspace{5mm}
{ \begin{tabular}{|l|c|c|}
\hline 
Grid&$E^{*}_{3}$,\AA$^{-2}\times 10^{-5}$&$E_{2}$,\AA$^{-2}\times 10^{-5}$\\
\hline 
\hline 
45$\times $45$\times $9&-22.819&-14.123\\
\hline 
60$\times $60$\times $9&-22.568&-13.913\\
\hline 
60$\times $60$\times $15&-22.570&-13.913\\
\hline 
75$\times $75$\times $9&-22.561&-13.907\\
\hline 
75$\times $75$\times $15&-22.563&-13.907\\
\hline 
90$\times $75$\times $9&-22.567&-13.912\\
\hline 
105$\times $75$\times $9&22.555&-13.902\\
\hline 
\end{tabular}\par}
\end{table} 
\begin{table}
\caption{Convergence of the He\protect$_{3}\protect$ ground state energy with respect
to the number of gridpoints}
\vspace{5mm}
{ \begin{tabular}{|l|c|c|}
\hline 
Grid&
$E_{3}$,\AA$^{-2}\times 10^{-5}$&
$E_{2}$,\AA$^{-2}\times 10^{-5}$\\
\hline 
\hline 
45$\times $45$\times $15&-1096.35&-13.839\\
\hline 
60$\times $60$\times $9&-1096.72&-13.894\\
\hline 
60$\times $60$\times $15&-1097.11&-13.894\\
\hline 
60$\times $60$\times $21&-1097.11&-13.894\\
\hline 
105$\times $60$\times $18&-1097.19&-13.9062\\
\hline 
105$\times $75$\times$15&-1097.25&-13.9062\\
\hline 
\end{tabular}\par}
\end{table} 
\begin{table}
\caption{Contribution of cluster wave to the Faddeev component}
\vspace{5mm}
{ \begin{tabular}{|c|c|c|}
\hline 
Potential&$\Vert f_{c}\Vert ^{2}$&$\Vert f_{c}^{*}\Vert ^{2}$\\
\hline 
\hline HFD-A&0.2094&0.9077\\
\hline HFD-B&0.2717&0.9432\\
\hline HFD-ID&0.1555&0.8537\\
\hline LM2M1&0.2412&0.9283\\
\hline LM2M2&0.2479&0.9319\\
\hline LM2M1a&0.2624&0.9390\\
\hline LM2M2a&0.2780&0.9458\\
\hline TTY&0.2487&0.9323\\
\hline 
\end{tabular}\par}
\end{table} 
\begin{table}
\caption{The mean square radius of Helium molecules, \AA}
\vspace{5mm}
{ \begin{tabular}{|c|c|c|c|}
\hline 
Potential&Ground state of He$_{3}$&Excited state of He$_{3}$&He$_{2}$\\
\hline 
\hline HFD-A&6.46&66.25&88.18\\
\hline HFD-B&6.23&57.89&62.71\\
\hline HFD-ID&6.64&75.38&126.73\\
\hline LM2M1&6.35&61.74&73.54\\
\hline LM2M2&6.32&60.85&70.93\\
\hline LM2M1a&6.27&59.03&65.76\\
\hline LM2M2a&6.21&57.17&60.79\\
\hline TTYPT&6.33&60.81&70.70\\
\hline 
\end{tabular}\par}
\end{table} 
\begin{table}
\caption{The mean radius of Helium molecules, \AA}
\vspace{5mm}
{ \begin{tabular}{|c|c|c|c|}
\hline 
Potential&Ground state of He$_{3}$&Excited state of He$_{3}$&He$_{2}$\\
\hline 
\hline HFD-A&5.65&55.26&64.21\\
\hline HFD-B&5.48&48.33&46.18\\
\hline HFD-ID&5.80&62.75&91.50\\
\hline LM2M1&5.57&51.53&53.85\\
\hline LM2M2&5.55&50.79&52.00\\
\hline LM2M1a&5.51&49.28&48.34\\
\hline LM2M2a&5.46&47.72&44.82\\
\hline TTYPT&5.55&50.76&51.84\\
\hline 
\end{tabular}\par}
\end{table} 
\begin{table}
\caption{Comparison with the results of other authors}
\vspace{5mm}
{ \begin{tabular}{|c|c|c|c|c|c|}
\hline 
Observable&This work&\cite{Pandharipande}&\cite{Moto1}&\cite{Gloeckle}&\cite{Carbonell}\\
\hline \multicolumn{6}{|c|}{HFD-A(He)}\\
\hline $E_{3}$, K&-0.1171&-0.117&-0.114&-0.11&-0.107\\
\hline $E^*_{3}$, mK&
-1.665& &-1.74&-1.6&\\
\hline \multicolumn{6}{|c|}{HFD-B(He) }\\
\hline $E_{3}$, K&-0.1330& &-0.132& &-0.130\\
\hline $E_{3}^*$, mK &-2.734& &-2.83& & \\
\hline 
\end{tabular}\par}
\vspace{5mm}
{ \begin{tabular}{|c|c|c|c|}
\hline 
\multicolumn{4}{|c|}{LM2M2 }\\
\hline 
\hline Observable&This work&\cite{Fedorov}&\cite{PRL1999}\\
\hline 
\hline $E_{3}$, K&-0.1264&-0.1252&-0.219\\
\hline $E^*_{3}$, mK&-2.271&-2.269&-1.73\\
\hline $<r^{2}>^{1/2}$, \AA&6.32&6.24&7.4\\
\hline $<r^{2}_{*}>^{1/2}$, \AA&60.85&60.86&50.3\\
\hline 
\end{tabular}\par}
\end{table} 
\begin{table}
\caption{Contribution of different two-body angular states to the Faddeev component}
\vspace{5mm}
{ \begin{tabular}{|c|c|c|c|c|c|c|}
\hline 
&
\multicolumn{3}{|c|}{Ground state}&
\multicolumn{3}{|c|}{Excited state}\\
\hline Potential&S&D&G&S&D&G\\
\hline 
\hline HFD-A&0.9991043&0.0008859&0.0000095&0.9999964 &0.0000035&0.0000000\\
\hline HFD-B&0.9990000&0.0009890&0.0000107&0.9999952&0.0000048&0.0000001\\
\hline HFD-ID&0.9991709&0.0008200&0.0000088&0.9999972&0.0000028&0.0000000\\
\hline LM2M1&0.9990505&0.0009390&0.0000101&0.9999958&0.0000042&0.0000000 \\
\hline LM2M2&0.9990393&0.0009500&0.0000103&0.9999957&0.0000043&0.0000000\\
\hline LM2M1a&0.9990129&0.0009762&0.0000105&0.9999954 &0.0000046&0.0000001 \\
\hline LM2M2a&0.9989834&0.0010053&0.0000109&0.9999950&0.0000049&0.0000001\\
\hline TTY&0.9990332&0.0009561&0.0000104&0.9999956&0.0000043&0.0000000\\
\hline 
\end{tabular}\par}
\end{table} 
\begin{table}
\caption{Contribution of different two-body angular states to the wave function}
\vspace{5mm}
{ \begin{tabular}{|c|c|c|c|c|c|c|}
\hline 
&
\multicolumn{3}{|c|}{Ground state}&
\multicolumn{3}{|c|}{Excited state}\\
\hline Potential&S&D&G&S&D&G\\
\hline 
\hline HFD-A&0.95416&0.03198&0.00877&0.90957&0.07543&0.01331\\
\hline HFD-B&0.95193&0.03365&0.00947&0.89710&0.08546&0.01546\\
\hline HFD-ID&0.95493&0.03116&0.00905&0.91919&0.06763&0.01170\\
\hline LM2M1&0.95323&0.03277&0.00891&0.90337&0.08043&0.01437\\
\hline LM2M2&0.95303&0.03294&0.00893&0.90201&0.08152&0.01460\\
\hline LM2M1a&0.95259&0.03332&0.00899&0.89904&0.08391&0.01512\\
\hline LM2M2a&0.95210&0.03374&0.00906&0.89574&0.08654&0.01569\\
\hline TTY&0.95245&0.03318&0.00941&0.90186&0.08164&0.01463\\
\hline 
\end{tabular}\par}
\end{table} 

\begin{figure}
{\centering \includegraphics{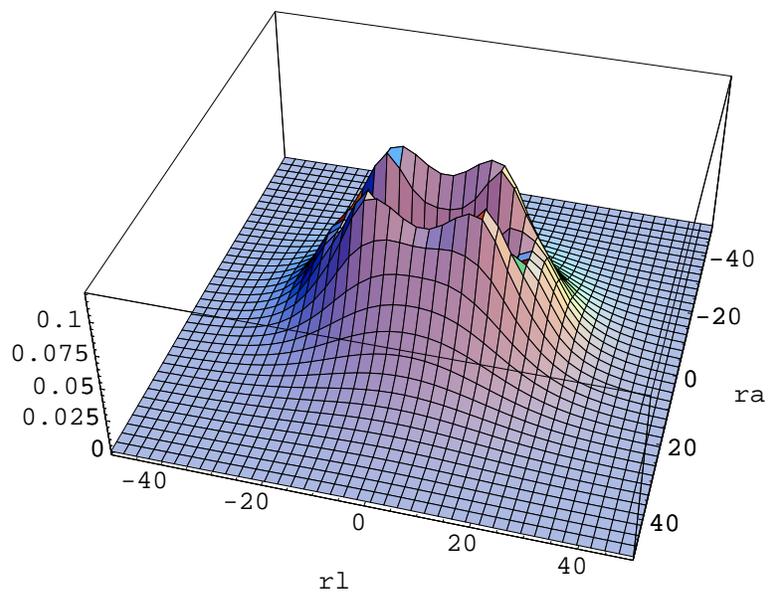} \par}
\caption{Conditional one-particle density function of the He$_3$ ground state, $r_l$, $r_a$ in \AA}
\end{figure}
\begin{figure}
{\centering \includegraphics{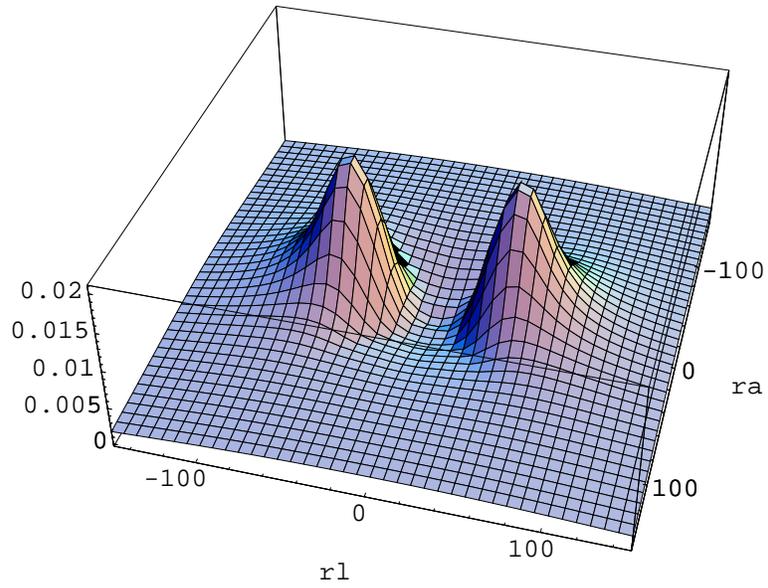} \par}
\caption{Conditional one-particle density function of the He$_3$ excited state, $r_l$, $r_a$ in \AA}
\end{figure}
\begin{figure}
{\centering \includegraphics{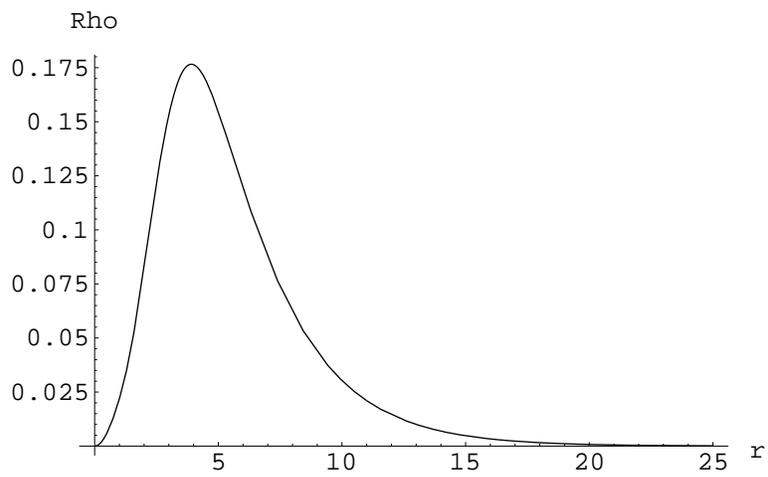} \par}
\caption{He$_3$ ground state density function, $r$ in \AA}
\end{figure}
\begin{figure}
{\centering \includegraphics{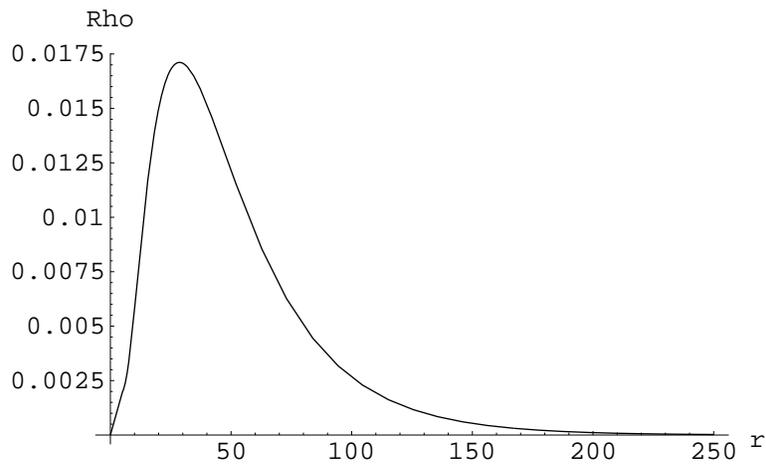} \par}
\caption{He$_3$ excited state density function, $r$ in \AA}
\end{figure}
\begin{figure}
{\centering \includegraphics{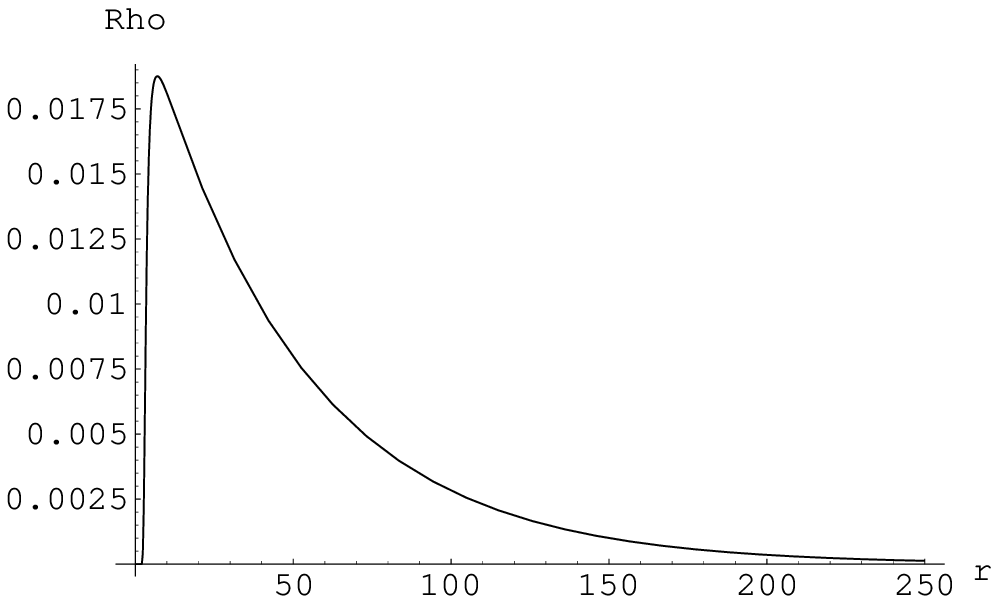} \par}
\caption{He$_2$ dimer density function, $r$ in \AA}
\end{figure}


\begin{thebibliography}{}
\bibitem{Experiment}W.Sch\"ollkopf and J.P.Toennies J. Chem. Phys. \textbf{104}(3), 1155, (1996)
\bibitem{Experiment2}F. Luo, C. F. Giese and W. R. Gentry J. Chem. Phys. \textbf{104}(3), 1151, (1996)
\bibitem{Experiment3}J. C. Mester, E. S.Meyer, M. W. Reynolds, T.E. Huber, Z. Zhao, B. Freedman,
J. Kim and I.F.Silvera Phys. Rev.Lett. \textbf{71}(9), 1343, (1993)
\bibitem{Efimov}V. Efimov, Phys.Lett. B \textbf{33}, 563 (1970)
\bibitem{Lim1}T. K. Lim and M.A.Zuniga J. Chem. Phys. \textbf{63}(5), 2245, (1974)
\bibitem{Lim2}T. K. Lim, S.K. Duffy and W.C.Damert Phys. Rev.Lett. \textbf{38}(7), 341, (1977)
\bibitem{Gloeckle}Th. Cornelius, W. Gl\"ockle, J. Chem. Phys., \textbf{85}, 3906 (1986)
\bibitem{Pandharipande}V. R. Pandharipande, J. G. Zabolitzky, S. C. Pieper, R. B. Wiringa, and U. Helmbrecht,
Phys. Rev. Lett., \textbf{50}, 1676 (1983). 
\bibitem{Moto1}E. A. Kolganova, A. K. Motovilov, S.A. Sofianos LANL E-print chem-ph/9612012 
\bibitem{Carbonell}J. Carbonell, C. Gignoux, S. P. Merkuriev, Few--Body Systems \textbf{15}, 15
(1993). 
\bibitem{Fedorov}E. Nielsen, D. V. Fedorov and A. S. Jensen, LANL e-print physics/9806020 
\bibitem{PRL1999}T. Gonzalez-Lezana, J.Rubayo-Soneira, S.Miret-artes, F.A. Gianturco, G. Delgado-Barrio
and P.Villarreal, Phys.Rev.Lett, 82(8), 1648, (1999)
\bibitem{CPC}V. Roudnev and S. Yakovlev, {\em Proceedings of the first international conference
 Modern Trends in Computetional Physics,} (1998), to be published in Comp. Phys. Comm.
\bibitem{FaddMerk}L. D. Faddeev, S. P. Merkuriev, \emph{Quantum scattering theory for several
particle systems} (Doderecht: Kluwer Academic Publishers, (1993)). 
\bibitem{TAM}V. V. Kostrykin,~A.~A.~Kvitsinsky,S.~P.~Merkuriev~Few-Body~Systems, \textbf{6},
97, (1989) 
\bibitem{Groning}N. W. Schellingerhout, L. P. Kok, G. D. Bosveld ~Phys. Rev. A \textbf{40}, 5568-5576,
(1989) 
\bibitem{LM}B. Liu and A. D. McLean, J. Chem. Phys. \textbf{91}(4), 2348 (1989)
\bibitem{Az79}R. A. Aziz, V. P. S. Nain, J. S. Carley, W. L. Taylor, and G. T. McConville,
J. Chem. Phys. \textbf{70}, 4330 (1979). 
\bibitem{Az87}R. A. Aziz, F. R. W. McCourt, and C. C. K. Wong, Mol. Phys. \textbf{61}, 1487
(1987). 
\bibitem{Az91}R. A. Aziz and M. J. Slaman, J. Chem. Phys. \textbf{94}, 8047 (1991). 
\bibitem{VanMourik}T. van Mourik and J. H. van Lenthe, J. Chem. Phys. \textbf{102}(19), 7479 (1995) 
\bibitem{TTY}K.T.Tang, J. P. Toennis and C. L. Yiu Phys. Rev.Lett. \textbf{74}(9), 1546,
(1995)
\bibitem{MonteCarlo}J. B. Anderson, C.A. Traynor and B. M. Boghosian, J. Chem. Phys. \textbf{99}(1),345
(1993) 
\end{thebibliography}
\end{document}